# Direct Printing in Three-Dimensions of 2D Material Inks

C. Grotta[1+], P. Sherrell[1+], M. Sokolikova[1], P. Palczynski[1], K. Sharda[1], A. Panagiotopoulos, C. Mattevi[1]*

[1.] Department of Materials, Imperial College London

* c.mattevi@imperial.ac.uk

+ equally contributed to this work.

**Abstract**

Deterministic assembly of 2D materials in three-dimensional miniaturized structures is paramount for the fabrication of new small-scale devices to serve in on-chip technologies. This is a new manufacturing challenge, as miniaturization is currently developed to achieve planar-geometries. Three-dimensional printing allows templating materials with arbitrary complexity at different scales, however, the available printable formulations are still limited to ceramics, polymers, metals and biological materials and none is suitable for 2D materials. Here, we show a general strategy to formulate highly concentrated, water-based 2D material inks with thermoresponsive characteristics enabling additive manufacturability of arbitrary self-supporting 3D architectures. The printed structures are extended over $mm^2$ in three-dimensions with fine micron sized features. Their mechanical robustness and chemical stability enable their employment as miniaturized hybrid supercapacitor-battery-like devices. The unique microstructure generated by the 2D crystals leads to demonstrate exceptionally high energy density of 0.5 mW h/$cm^2$.

**Main Text**

Miniaturized electrochemical energy devices are crucial for on-chip technologies [1], these require miniaturized three-dimensional structuring to ensure sufficient material utilization to obtain desirable energy densities [2-4] over small footprint areas. Direct-ink writing, or robocasting, is the technique of choice for printing materials from an ink gel into any arbitrary shape. The final architecture is formed by depositing a continuous filament via extrusion of an ink-gel in a layer-by-layer fashion [5]. 2D crystals of the class of transition metal dichalcogenides have been recently recognized as exceptionally promising materials for energy storage devices [6].

Printing of 2D crystals in three-dimensions requires fundamentally new formulations which are very different from the ones known for printing over two-dimensions (ink-jet printing) [7]. Structural materials requirements become critical to obtain free-standing mechanically stable architectures extended over three-dimensions with micrometric features. These impose ultra-high concentration of the 2D crystals in the inks and large sizes of crystals with atomically thin thickness [8]. Further, the inks are required to flow through the deposition nozzle at high shear stresses and to set immediately once out from the nozzle [5,9] in order to achieve shape retention. This behaviour is called shear-thinning and it can be shown by fluids with viscoplastic properties. The viscoplasticity requires establishing good non-covalent interactions between the different ink components to ensure fluency and, at the same time, cohesiveness. This is particularly challenging for 2D van der Waals crystals due to their inherent chemical inertness [10, 11] which hinders the formation of these good non-covalent interactions as well as renders covalent functionalization arduous. To circumvent this challenge, it is crucial to utilize large mono- and few- layered sheets of these 2D crystals as they present strong liquid-like adhesion energies which are not found in their bulk form [12]. Mono- and few- layered sheets with large lateral size (>1 µm) are also essential to ensure mechanical integrity and stability of the 3D architecture [13].

Here, we have developed non-hazardous water-based inks of exfoliated 2D crystals by exploiting the amphiphilic properties of a thermoresponsive-triblock polymer (poloaxomer). The poloaxomer anchors the hydrophobic sheets, preventing restacking, and allows for water dispersibility. This ensures the individually exfoliated sheets are physically distinct during the printing procedure. The thermoresponsivness enables controlled gelation, which is necessary to retain the imparted shape during the printing process. We demonstrate the feasibility of this process in printing, *via* robocasting, 3D architectures with arbitrary complexity of different 2D crystals (Figure 1).

The fabrication process of 3D-printed architectures is illustrated in Figure 1. Highly concentrated inks have been achieved via exfoliation of lithiated transition metal dichalcogenides (TMDs) sheets in concentrated poloaxomer solution. The poloaxomer of choice consists of a poly(propylene oxide) block between two poly(ethylene oxide) blocks (Figure S1a), called Pluronic F127, is non-hazardous, biocompatible and commercially available. The pre-lithiation ensures effective exfoliation occurring in contact with the water-based poloaxomer solution which leads to single- and few-layered TMD sheets with a lateral size in the micrometre range and high monodispersivity [1]. In addition to electrostatic stabilization, the amphiphilic polymer provides a secondary steric stabilization factor which significantly increases retention of mono- and few-layer exfoliated flakes in solution by providing a physical barrier to restacking [14]. The ink exhibits shear-thinning behaviour and thus is directly extruded from a nozzle to form 3D architectures [5]. We demonstrate printing of exfoliated $MoS_2$ and $TiS_2$ 3D structures and we show device implementation using $MoS_2$. Traditional methods of lithium-intercalation and exfoliation result in TMD concentrations below 1 mg/mL, which is too low to form mechanically stable structures, to meet the required rheological properties for printing, and to have sufficient active material for excellent device performance [15]. Our poloaxomer-assisted exfoliation method overcomes this limitation by functionalizing, stabilizing, and concentrating exfoliated TMDs during exfoliation. (Figure 1, Figure S1b).

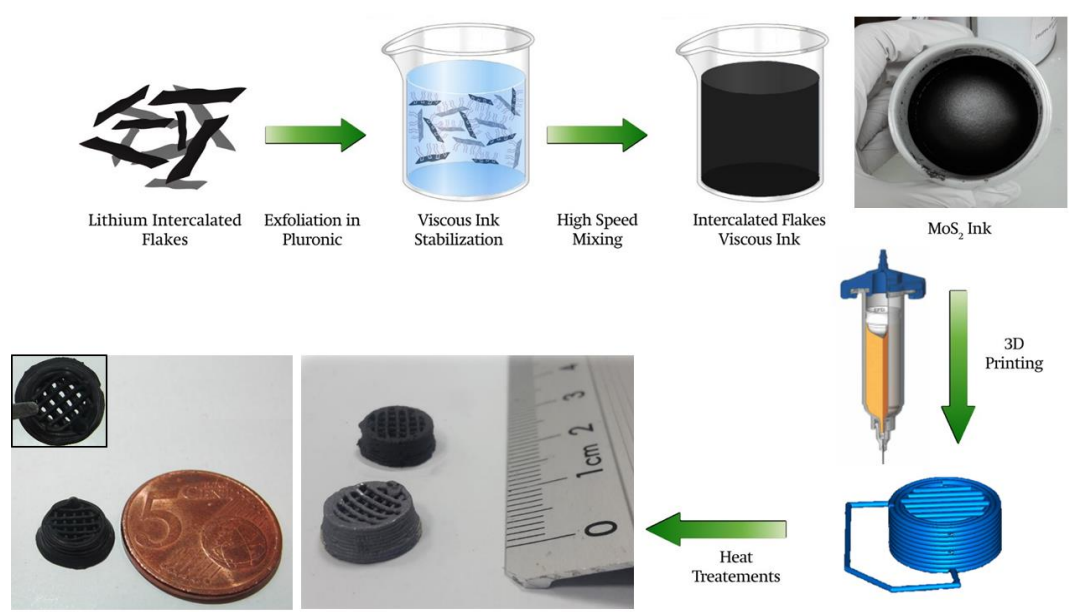

*Figure 1. Schematic showing the exfoliation process for the production of a 2D crystals ink from bulk crystals through to robocasted 3D printed architectures with a wood-pile structure (blue-grey $TiS_2$, dark grey $MoS_2$).*

The higher viscosity ($10^2$ mPa s) of the poloaxomer solution with a concentration of 25% w/w (Figure S2) compared to DI water (0.89 mPa s) causes entrapment of produced gas bubbles in the solution. In order to completely exfoliate our TMDs in this viscous solution a homogenization step and degassing step is implemented, which mechanically removes the gas bubbles in solution and allows the water molecules to fully interact with the Li-TMD powder (Figure S1c).

TEM analysis of $MoS_2$ inks has demonstrated that the inks contain mono- layered sheets

(Figure 2a-d) with lateral sizes of ~ 1 μm (Figure 2a-c). Moire patterns arising as a result of angular rotation of superimposed single-layered $MoS_2$ nanosheets can indicate the tendency of an exfoliated material to restack partially (Figure 2d). XRD analysis has further confirmed exfoliation of $MoS_2$ with the poloaxomer (Figure S3a). The inks are formed by both 1T and 2H $MoS_2$ phases. A 1T component is expected as we are using the Lithium-intercalation-exfoliation method [16] (Figure S3b).

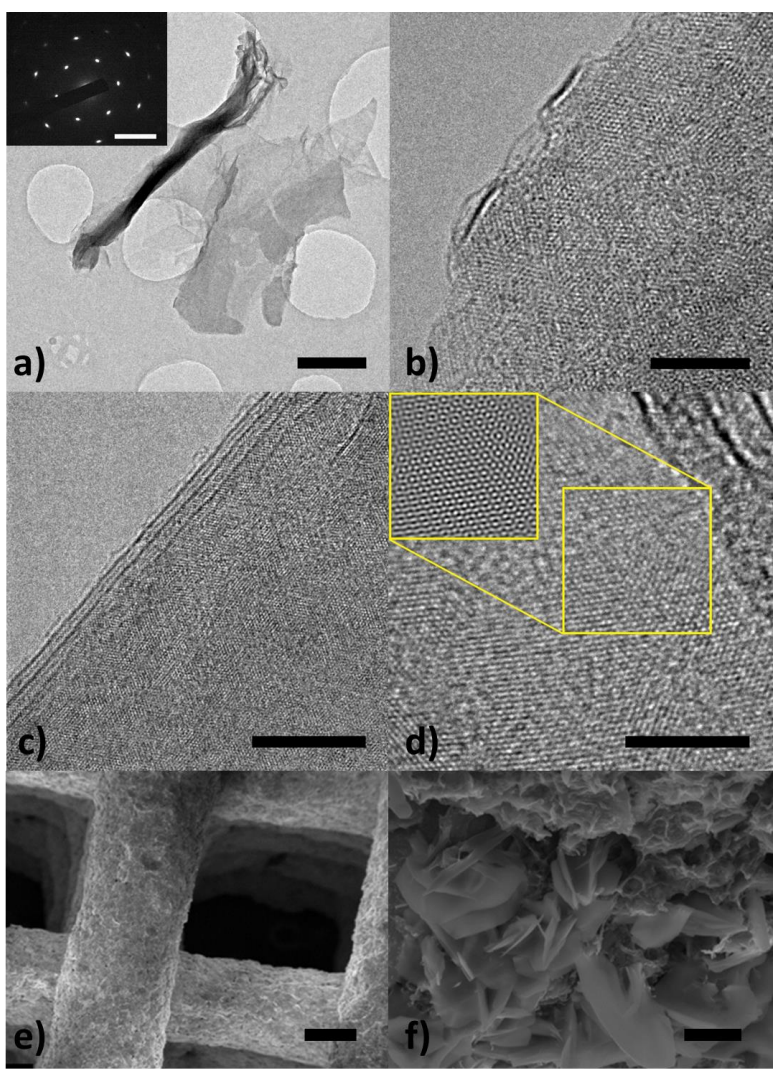

*Figure 2. Structural characterization of the exfoliated $MoS_2$ nanosheets and the 3D architectures: Electron micrographs of poloaxomer-LI exfoliated $MoS_2$ nanosheets; a) TEM image of exfoliated $MoS_2$ nanosheets (scale bar = 200 nm) with SAED pattern (inset, scale bar = 5 $nm^{-1}$); b) high-resolution TEM images of the edges of $MoS_2$ nanosheets from the 2D crystal ink showing a single-layered sheet (scale bar = 5 nm); c) TEM of three-layered $MoS_2$ nanosheets (scale bar = 5 nm); d) TEM image and a Fourier filtered image (inset) highlighting a Moire pattern formed upon restacking of single-layered $MoS_2$ nanosheets*

*(scale bar = 5 nm);* **SEM images of an annealed 3D printed MoS$_2$ architecture**; *e) struts of a wood-pile architecture; and f) struts surface morphology with visible MoS$_2$ nanosheets. (scale bars, e= 300μm, f= 10μm).*

When the initial vigorous reaction has subsided, a high-speed mixing step is introduced (Figure 1, S1c) to homogenise and degas the exfoliated MoS$_2$ ink. The amphiphilic nature of the poloaxomer allows for stabilization of concentrations of MoS$_2$ sheets on the order of 10 mg/mL. To convert the liquid inks into a solid structure, the ink must flow through the nozzle and gel upon completion of the extrusion process. The gelation temperature of ~ 25 °C of Pluronic F127 is thus necessary to enable shape retention upon printing at room temperature. By tuning the MoS$_2$ concentration, the viscoelastic properties required for 3D printing have been achieved (Figure S2) with exfoliated MoS$_2$ nanosheets in a molecular weight ratio of 40% with respect to pluronic. Inks with lower MoS$_2$ contents (between 25% and the 30% wt.) present a viscosity, which is too low to enable the shape retention of the printed architecture (Figure S3). Conversely, increasing the concentration of MoS$_2$ up to 60% wt. led to inks with viscosity too high to enable the flow through the nozzle.

The exfoliated MoS$_2$ 40% wt. ink is the optimum formulation to ensure fluidity of the ink and stiffness of the printed structure. Under rheological investigation, these inks show shear thinning behaviour and viscosity in the range of $10^2$ to $10^4$ mPa s, which confirms their suitability for robocasting [17-19]. Our inks require different viscoplasitc properties in comparison to reported ink jet formulations where a low surface tension and viscosity on the order of 1 mPa s is required [20, 21]. The ink jet printed formulations based on triton x-100 and propylene glycol are able to achieve 2D percolation at concentrations between 2-3 mg/mL, however, significantly higher concentrations are required to achieve percolation in 3D. Our inks demonstrates such 3D percolation at a concentration 10 mg/mL upon printing due to our use of the lithium exfoliation strategy resulting in significantly larger lateral size stable monolayer flakes compared to the sonochemical or shear exfoliation strategies described elsewhere (lateral size < 5 μm vs < 400 nm respectively) [20, 21]. A comparison of viscosity for different printable inks and the relative mass loading for jet printing and 3D robocasting formulation based on 2D materials, nanomaterials and ceramic particles is reported in Figure 3. While 2D materials inks for jet printing and graphene inks for 3D printing show mass loading comparable with the current state of the art suspension of these materials, our inks based on TMDs show (Figure 3) a few order of magnitude higher mass loading and comparable to ceramic particle inks. As the volume fraction of material required for percolation is inversely proportional to flake diameter [22] our larger lateral size flakes achieve 3D percolation at concentrations significantly lower than sonochemically exfoliated flakes, albeit at higher concentrations than comparable 1D materials [23].

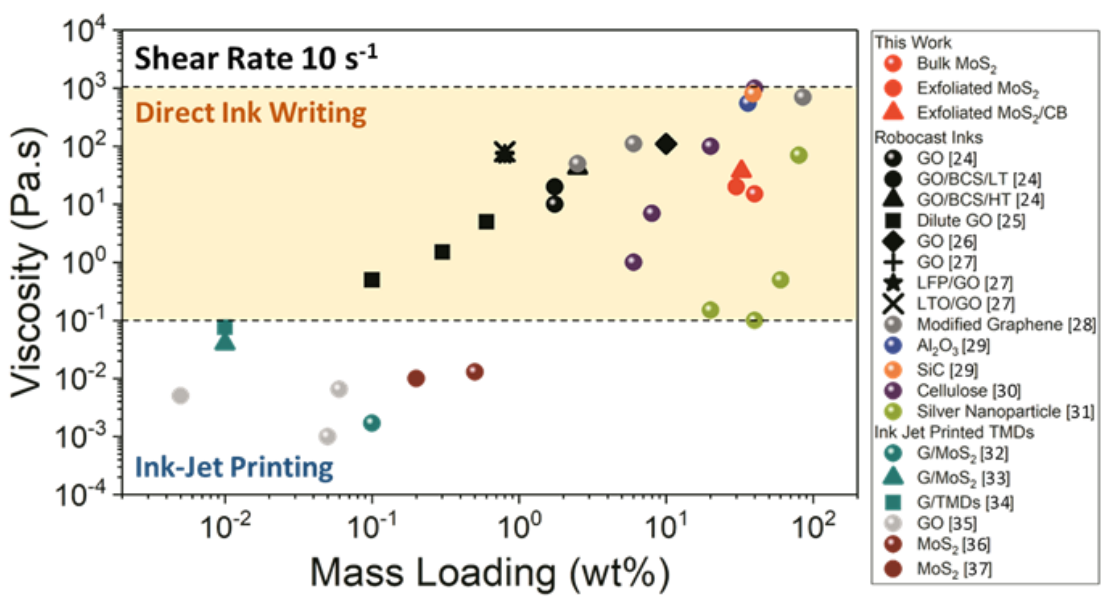

*Figure 3. Comparison of viscosity and mass loading for robocast inks and ink jet printed 2D materials. Demonstrating the potential of our polymer-assisted exfoliation to attain high mass loadings of 2D materials.*

Due to percolating network from large lateral size, monolayer TMDs, the printed architectures are mechanically stable and show a good adhesion layer-to-layer with no deformation (Figure 1). The extruded filaments are uniform (Figure 2e) and they appear as a dense network of evenly distributed $MoS_2$ sheets (Figure 2f, g). After printing, the architectures were dried in air for 2 days and subsequently annealed at 600°C in an argon atmosphere for up to 2 hours to remove the resistive poloaxomer. As a result of this thermal treatment, there is a full conversion of the 1T phase of $MoS_2$ into the 2H phase [16]. The effect of the removal of the poloaxomer via annealing is to decrease the electrical resistance from 15 kΩ/sq to ~1 kΩ/sq whilst increasing the available surface area (up to 70 $m^2$/g), and porosity (Figure S5, Table S1). The annealing leaves $MoS_2$ chemically pristine, as revealed by X-ray photoelectron spectroscopy (XPS) (Figure S6). Further, a drastic decrease of the carbon 1s peak indicates the effective removal of the pluronic. For comparison, inks of bulk crystals of $MoS_2$ and poloaxomer were prepared, by blending the two in a high-speed mixer (Figure S1b). The mechanical stability of the printed architectures post annealing is attributed to the highly exfoliated $MoS_2$ sheets. The high concentration (up to 10 g/L) of these highly stable, well exfoliated $MoS_2$ sheets allows the formation of a load bearing network throughout the printed architectures. This load bearing network is achieved as lowering the aspect ratio of the $MoS_2$ along the c-axis from bulk to monolayer decreases dramatically the percolation threshold concentration.

The exfoliated–$MoS_2$ structures are mechanically strong, exhibiting compressive stress of 0.2 MPa (Figure S7) which is comparable to rGO printed architectures previously reported [37, 38]. While the $MoS_2$-bulk structure show poorer mechanical strength, with a compressive stress of 0.1 MPa, emphasising the importance to have well exfoliated 2D crystals, which can form strong adhesion forces to produce mechanically strong and flexible struts. The annealing process led to crumbling, and soaking of the as produced $MoS_2$-bulk architectures resulted in slow dissolution of the structure due to a lack of entanglement between the $MoS_2$ sheets. This crumbling of the $MoS_2$-bulk architectures post-annealing is strong evidence for the mechanical strength of the $MoS_2$-exfoliated architectures arising from a percolating, self-supporting network of exfoliated sheets.

Due to the flexibility of the ink preparation method, functional nanomaterials can be easily incorporated to tailor properties (Figure S8). Carbon black has been added into the exfoliation at the weight percentage of 2.5% to tailor the porosity and electrical properties of the architectures. Printed architectures of exfoliated $MoS_2$ (30% wt.) incorporating 2.5% wt. carbon black have demonstrated an enhancement of the specific surface area, from 16 $m^2$/g to 70 $m^2$/g as compared to pure $MoS_2$ structures along with the number of pores with radius under 1 nm (Table S1).

The high $MoS_2$ edge density, low electrical resistivity, mechanical strength, and pore structure make these $MoS_2$/CB architectures promising for energy storage, thus the electrochemical performance of the 3D printed architectures was investigated. Electrode configurations for printed, miniaturized hybrid electrochemical energy devices fall into two general categories: interdigitated and stacked. In the former, the electrode structures are complementary in shape, while in the latter two similarly structured electrodes are separated by an ion permeable membrane and overlayed with the structure [1]. Due to the versatility of the produced inks, both configurations were demonstrated (Figure 4a, b). Further, they show stability and adhesion onto rigid and flexible substrates (Figure 4b). This ability to design and print custom geometries on arbitrary substrates is a powerful tool for building electrochemical devices for applications requiring specific electrode shapes and sizes.

Device testing was performed on stacked electrode configurations (interdigitated electrode results shown in Figure S9), as such an architecture maximised the materials volume per unit area. To determine the power and energy density of our 3D structures, we studied their performance in both aqueous and non-aqueous electrolytes. Specifically, we tested their properties in both $H_2SO_4$ in DI water and tetraethylammonium tetrafluoroborate (TEA $BF_4$) in propylene carbonate electrolytes to probe the materials performance in both a small ionic radius electrolyte [$H_2SO_4$], which can potentially lead to a high power density, and a large potential window electrolyte [TEA $BF_4$] that is comparable to that used in commercial devices.

Electrochemical characterization via cyclic voltammograms and galvanostatic charge/discharge of $MoS_2$ structures with integrated CB were performed. From both cyclic voltammetry (Figure 4b,c) and galvanostatic charge/discharge (Figure S10) the maximum areal capacitance of annealed $MoS_2$/CB stacked electrode configurations was determined to be 900 mF/cm$^2$ in $H_2SO_4$ and 1450 mF/cm$^2$ in TEA $BF_4$ (Figure 4c-e). The cyclic voltammograms (Figure 4c) of stacked structures show a rectangular shape in $H_2SO_4$ indicating a capacitive behaviour of the architecture. The electrodes demonstrate some charge transfer and diffusion limitation only allowing for access to the maximum areal capacitive values below 10 mV/s (Figure 4e).

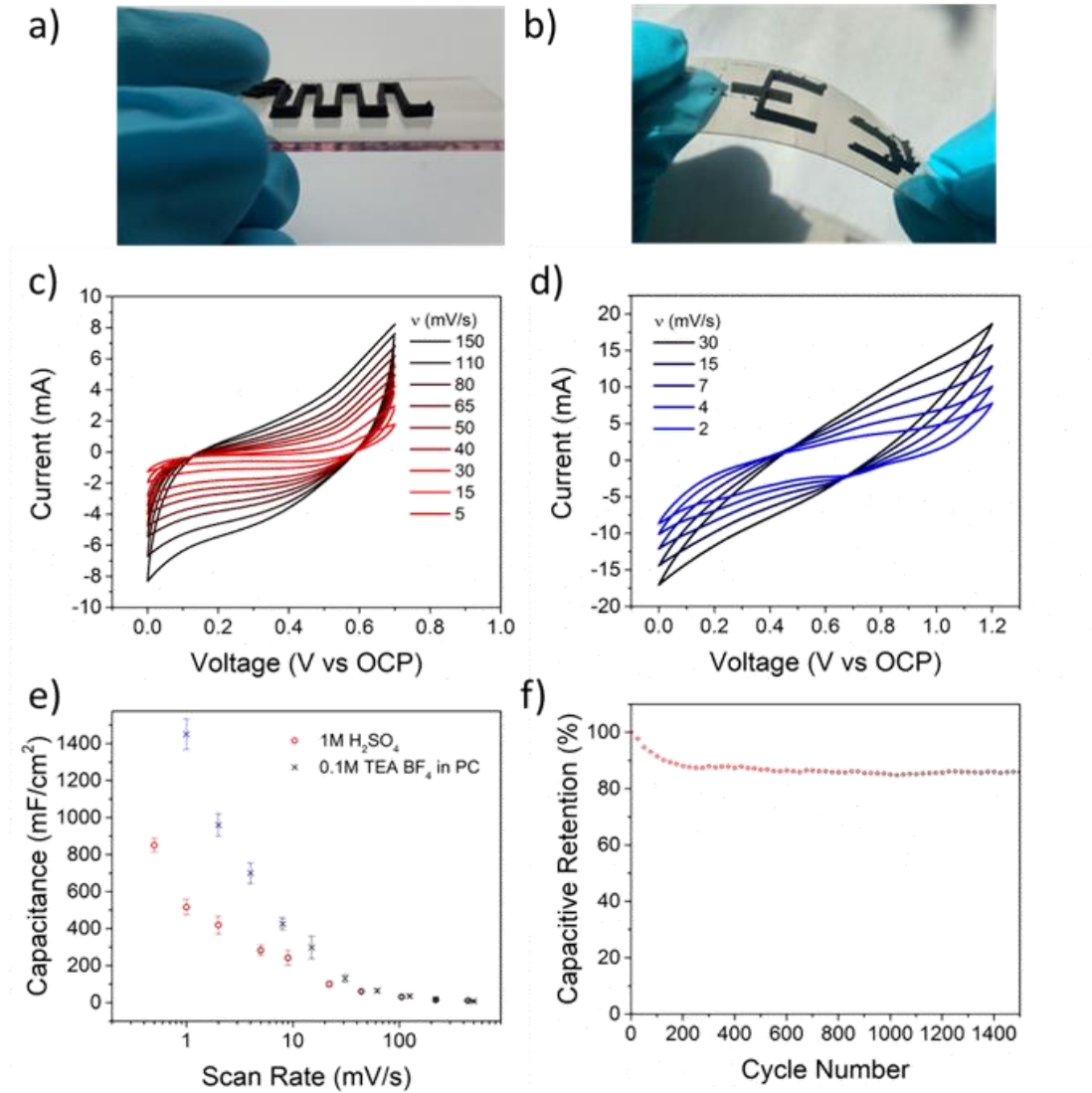

**Figure 4. Electrochemical characterization of the stacked MoS₂ architectures**: *a) Photographs of 3D printed stacked electrode architecture; b) 3D printed interdigitated electrodes on a flexible substrate; c) cyclic voltammogram of a stacked electrode architecture of MoS$_2$/CB on glass slides in 1M H$_2$SO$_4$; d) in 0.1M TEA BF$_4$ in propylene carbonate; e) effect of scan rate on capacitance for stacked electrode architectures of MoS$_2$/CB in 1M H$_2$SO$_4$ and 0.1M TEA BF$_4$/PC showing an areal capacitance of 1450 mF/cm$^2$ for 0.1M TEA BF$_4$/PC; f) capacitive stability over 1500 cycles at 10 mV/s for a stacked architecture in 1M H$_2$SO$_4$.*

In TEA BF$_4$ (Figure 4d) there is a significant increase in current when compared to H$_2$SO$_4$ at equivalent scan rates below 30 mV/s, however, the slow diffusion of the TEA ions into the meso-pores of the architecture limits the achievement of higher currents at faster scan rates. This difference is best reflected by comparing the RC time constant, τ, of the H$_2$SO$_4$ and TEA $BF_4$ where the values starkly contrast as 4.3 s and 108 s respectively. Electrochemical impedance spectroscopy (EIS) supports the slow time constants and high areal capacitance of the device configurations (Figure S11). The EIS shows a negative phase peak at $10^{-1}$ and $10^{-2}$ for $H_2SO_4$ and TEA $BF_4$ respectively, providing a τ of 10s and 100s which corroborates the CV data. Normalizing the real capacitive component extracted from the EIS data for geometric area also shows a maximum capacitance of 0.16 and 0.8 F/g for $H_2SO_4$ and TEA $BF_4$ respectively at a frequency of 1 mHz, however there is no observed capacitive plateau indicating that this capacitance will continue to increase at a slower frequency regime, as we observed in the cyclic voltammograms. Overall, the EIS supports the achievement of a high energy density material that is rate limited, which arises due to the high density of edge sites for charge generation and storage along with 100s of micron strut sizes. We envisage that decreasing this strut size will allow access to the charge generated at these edge sites at significantly faster rates, leading to an increase in the power density of our devices.

The $MoS_2$/CB annealed stacked architecture shows cyclic stability over 1500 cycles at a current density of 12 mA/cm$^2$ (Figure 4f) of ≈ 90% of the initial capacitance retained. The slight decrease in capacitance over the first few cycles arises from a decrease in surface area due to initial assessment of the electrode architecture in the electrolyte; once these surface defects are removed the remaining material demonstrates stability. A highly stable capacitance of 900 mF/cm$^2$ is exceptionally exciting for 3D printed structures, comparing favourably to those reported for other 3D printed materials for miniaturized supercapacitors (Table S2).

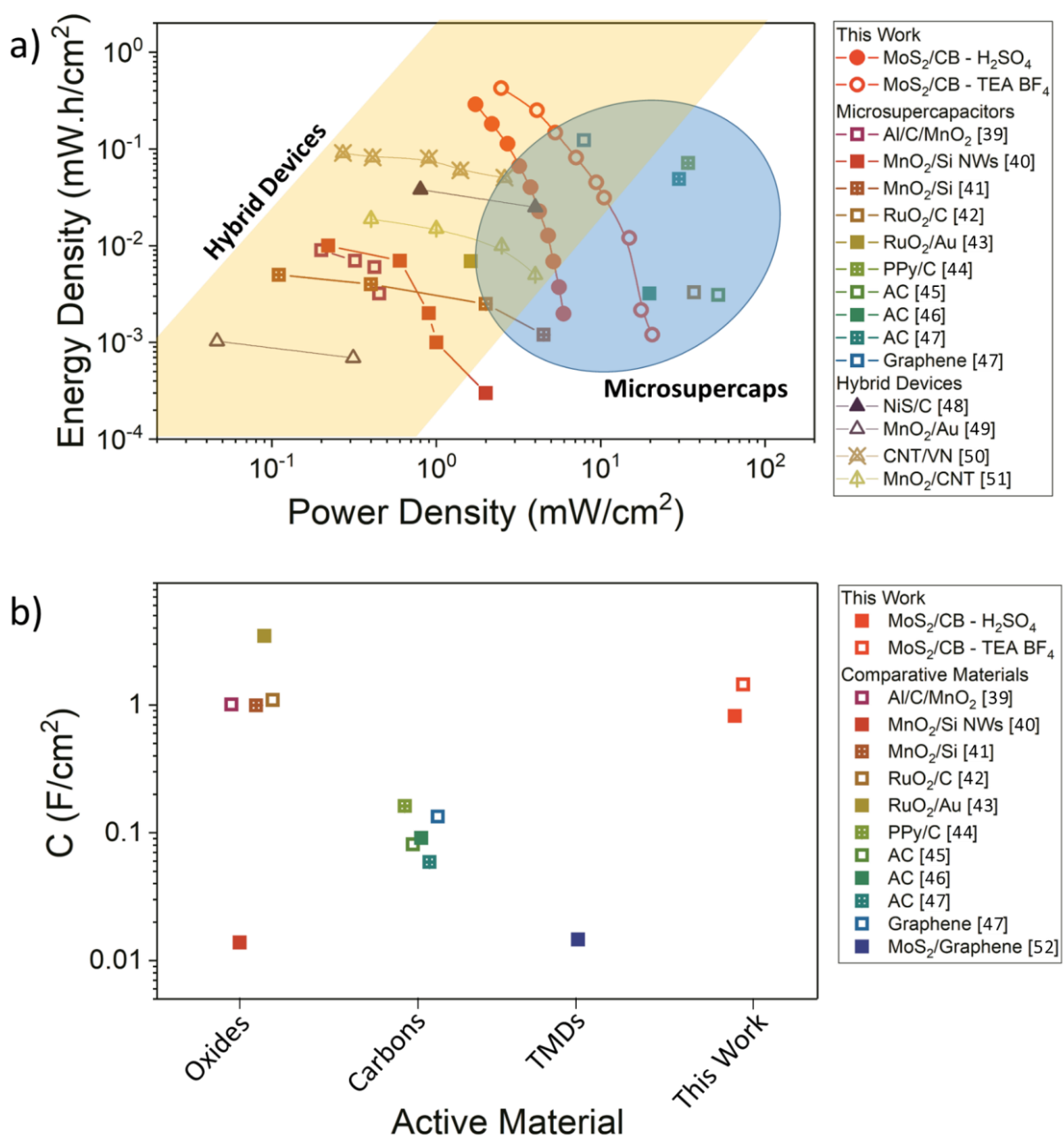

**Figure 5. Electrochemical performance**: *a) Ragone plot of the areal energy and power densities of 3D MoS$_2$ miniaturized electrochemical devices (microsupercapacitors, hybrid devices) in comparison to literature; b) areal capacitive comparison for microsupercapacitors fabricated from various oxide, carbon, and TMD materials.*

An effective way to characterise device performance under a variety of different conditions is through a Ragone plot (Figure 5). This plot shows a clear relation between the design of the TMD architectures and their performances, with the multiple plotted values arising from device performance under different testing conditions, in our case corresponding to the applied voltage scan rate. The increased surface area and micro- porosity of exfoliated $MoS_2$/CB architectures increases the power density by a factor of 3 and the energy density of nearly two orders of magnitude (Figure 5). The exceptional energy density is proven to arise from the highly exposed $MoS_2$ edge facets, rather than from the addition of CB, by comparing our free-standing $MoS_2$/CB architectures with free-standing CB only architectures (Figure S12), which have an equivalent τ of 12.5s yet a capacitance 4 orders of magnitude greater. An additional evidence that the exfoliated $MoS_2$ edges critically contribute to our high energy density is provided by the lower energy density of the $MoS_2$ bulk structures at equivalent mass loadings (0.5 mW h/cm$^2$ – $MoS_2$/CB; 0.02μW h/cm$^2$ – bulk $MoS_2$).

The subsequently increased surface area, decreased resistivity (250 Ω.cm), and changes in pore structure by the addition of carbon black further increase the performance of the devices, resulting in the highest attained energy density of 0.5 mW h/cm$^2$ and power density of 20 mW/cm$^2$ for full stacked configuration device architectures (Figure 5a). These characteristics, specifically the exceptional energy density and high capacitance, demonstrate our printing approach and material is ground-breaking in the emerging hybrid supercapacitor-battery device field [40]. The prime areal energy density attained of 0.5 mW h/cm$^2$ arises due to the high availability of $MoS_2$ edge facets in our printed struts along with the high proportion of pores under 1 nm. It is noteworthy that such energy density would be sufficient to power small devices such as implantable biomedical devices, environmental sensors and microelectromechanical systems [41].

When comparing our devices to microsupercapacitors (Figure 5) based on rGO/carbon nanotube, they out-perform with significantly higher energy density (0.4 mW h/cm$^2$ to 0.016 mW h/cm$^2$) (Figure 5a), areal capacitance (1450 mF/cm$^2$ compared to 116.3 mF/cm$^2$) (Figure 4b) and power density (20 mW/cm$^2$ compared to 2.84 mW/cm$^2$) [55]. The best performing microsupercapacitors demonstrated so far, present a combination of oxide nanoparticles with conductive materials such as carbon [28, 56, 57] or silicon nanowires [40]. While these materials compositions lead to significantly superior power densities, the energy densities compared to carbon-based devices, their performances are either lower or comparable to our $MoS_2$/CB architectures. In comparison to hybrid supercapacitor-battery devices (Figure 5a) our material out-performs the best reported CNT-vanadium nitride device in terms of maximum power, and energy density, whilst moving from a fibre style geometry to a full 3D printed architecture [50]. As it appears in the Ragone plot, the high rate of energy density increases by lowering the power density, indicating that our devices are operating in a kinetically limited regime where the combination of charge transfer and diffusion limitations are inhibiting rapid release of the stored energy. This suggests that by increasing the electrical conductivity of our inks further, significant gains in power density may be attainable. The versatility of our ink preparation methods would allow for the addition of functional additives, such as metal nanoparticles, carbon nanotubes, or graphene, to boost conductivity and increase pore accessibility further. A critical figure of merit for microsupercapacitors is also the areal capacitance [1]. Our $MoS_2$/CB structures show enhanced values compared to non-oxide based microsupercapacitors (Table S2) based on planar geometries.

To conclude, we demonstrate the formulation of 3D printable inks of chemically inert 2D crystals to template 3D miniaturized architectures of arbitrary complexity. The inks formulation and printing conditions of 2D crystals represent new processes in materials science and engineering which can be transferred to any other 3D printing robocasting, different 2D and nanomaterials in general. In this work, we show that the produced architectures can serve as miniaturized hybrid electrochemical devices for electronics in on-chip technologies. They are mechanically robust and chemically stable with features as small as 100 μm. They show leading areal capacitance and energy density as compared to the existing planar microsupercapacitors, and improved overall performance compared to hybrid supercapacitor-battery devices, paving the way towards additive manufacturability of 3D miniaturized devices.

**METHODS**

**Preparation of TMD Inks**

TMD powders (Alfa-Aesar) of either $MoS_2$ or $TiS_2$, and $LiBH_4$ powders are mixed in the glove box and heated at 350 °C for 72 hours with a molar ratio of 1:2.5 (TMD:Li). The mixture is cooled down to room temperature linearly over 12 hours. The commercial poloaxomer Pluronic F-127 is used as received from Sigma Aldrich in granular state. 25wt% is dissolved in Millipore water and stirred in an ice bath for 5 hours. The intercalated $MoS_2$ is transferred into a fireproof plastic container, model THI150ML 150ml HDPE thick wall jar with inner lid & cap, dispensed by THINKY©. The prepared gel is then poured to the mixture to get mixing the two components together energetically with a steel spatula.

Concentrations of up to 60 wt% TMD have been prepared. To ensure complete homogeneity, the inks were mixed using a planetary mixer, (Model THINKY ARE-250 Mixing and Degassing Machine, Conditioning Planetary Mixer). Immediately after exfoliation the ink appeared fragmented. The first mixing at high speed helps to unify the slurry.

**Rheology of MoS$_2$ Inks**

Inks rheology was tested using a Discovery Hybrid Rheometer HR1 (TA Instruments). The inks were tested using a parallel plate-plate of 40 mm diameter, in a range of shear rate from 0.1 to 1000 s$^{-1}$. Measurements were performed at a fixed room temperature of 25°C using a gap of 1000 μm.

**Printing TMD Inks**

The ink was transferred into a syringe for extrusion (model Optimum Syringe Barrels 30cc from Nordson EFD). Nozzle diameters to shape out 3D structures of 0.58, 0.20 and 0.10 mm were found to be suitable (model 20 Pink 0.58, from Nordson EFD). The syringe is fixed into the robosystem 3D printer at which a needle is attached. The strength at which the paste is extruded was chosen at an intermediate pressure of 10 mm/s. At this pressure, in accordance with the viscosity of the ink, the diameter of the needle is important to fit the process. All structures were printed on a Teflon substrate for easy removal. The printing speed was set at 10 mm/s. The architectures are built through a layer by layer deposition. TGA was used to estimate the optimal temperature to effectively burn out the polymer and avoid any degradation of the TMD crsytals.

**Ink Dilution**

Ink dilution for material characterization was performed simply by extruding a small amount of exfoliated MoS$_2$ ink into a vial followed by the addition of DI water. The vial was sealed and manually shaken to obtain a pseudo-stable dispersion which could be analysed over 48 hours before precipitation. The diluted inks were used for TEM, XRD, UV-Vis absorption spectroscopy, and XPS analyses.

**Thin Film Preparation**

Thin films were prepared from diluted ink solutions using a space confined interfacial method previously developed in our group [3]. 10 mL of diluted ink solution, 200 μL of HCl, and 1mL of Hexane were mixed by manual shaking. A cleaned 200 nm Si/SiO$_2$ substrate was lowered at 5 mm/min into the solution and held for 5 minutes. Finally, the substrate was raised at 1 mm/min from the solution before being allowed to air dry for 5 minutes. This process was repeated between 1 and 10 times to produce homogenous films for XRD, UV-Vis absorption spectroscopy, and XPS characterization.

**Transmission Electron Microscopy**

Samples for TEM analysis were prepared by drop-casting diluted dispersions of MoS$_2$ inks on carbon-coated copper grids. The grids were dried overnight under ambient conditions. Low-magnification bright-field TEM images and selected area electron diffraction (SAED) patterns were acquired on a JEOL JEM-2100Plus microscope. High-resolution TEM (HRTEM) imaging was performed using a JEOL JEM-2100F microscope with a field-emission gun operated at 200kV accelerating voltage.

**X-Ray Diffraction**

XRD patterns of MoS$_2$ films were studied via a Panalytical X'Pert Pro diffractometer. Scans were performed from 5 - 80° 2θ with a step size of 0.033°. X'Pert Highscore plus software was used for comparison with the database. High resolution scans over the (002) peaks were performed with a step size of 0.0085°.

**UV-Visible Spectroscopy**

UV-Vis was performed using a Lambda-25 UV-Vis spectrometer using 1 cm path-length cuvettes. Scans were performed between 1100 and 300 nm at 5 nm/s with a spectral resolution of 0.5 nm.

**Raman Spectroscopy**

Raman spectra were collected using a Renishaw inVia spectrometer equipped with a 532 nm laser. Maps were collected under a 100x objective with a grating of 1800 line/mm (spectral resolution 1.5 cm$^{-1}$).

**X-ray Photoelectron Spectroscopy**

XPS spectra were recorded on a Thermo Scientific K-Alpha$^+$ X-ray photoelectron spectrometer system. Data were collected at 20 eV pass energy for both core level (S 2$p$, C 1$s$, and Mo 3$d$) and valence band spectra and analysed using the Avantage software package.

**Scanning Electron Microscopy**

Scanning electron microscopy was performed on a Sigma300 LEO Gemini field-emission gun microscope equipped with a backscatter electron detector. The images were acquired in the range of 5-20 kV of accelerating voltage.

**Nitrogen Sorption**

Nitrogen sorption experiments were performed on a Quantachrome Autosorb iq system, data was analysed using the Quantacrhome ASiQWin software. Surface area was determined using the BET method. Pore size was determined based on DFT calculations built into the software.

**Mechanical Testing**

Compression cyclic tests were performed with universal testing machine Zwick/Roell Z010 with a loading cell of 10kN following the ASTM C133-97 standard. A metallic semi sphere was placed between the testing rig and the sample to ensure uniformity of the applied load. Cyclic tests were run at a rate of 1.5 mm/min imposing a maximum deformation limit of 5% of strain.

**Electrochemical Characterization**

The electrochemical measurements were performed in 3-electrode, 2-electrode, and device configurations using a potentiostat (Gamry Interface 1000, running Gamry Framework v6.25 Data Acquisition software). Cyclic voltammetry was performed at various electrochemical potential windows dependant on the electrolyte, at scan rates between 0.5 and 0.0001 V/s with a 1 mV step size. Galvanostatic charge/discharge tests were performed between symmetrical potentials (vs OCP) with a sampling period of 0.01 s and current densities ranging from 0.01 to 10 A/g.

**Free-Standing Wood-Pile Architecture Electrochemical Characterization**

Free-standing 3D architectures were adhered to conductive ITO glass substrates for ease of connection using LEI-C carbon cement. In both 3-electrode, and 2-electrode configurations a platinum mesh auxiliary electrode was used for charge compensation. The 3-electode system was used for 1M $H_2SO_4$ in DI water with a Ag/AgCl reference electrode (BaSI). The 2-electrode system was used for organic electrolytes, specifically TEA $BF_4$ in propylene carbonate, as these air sensitive experiments were performed in a glove box equipped with biaxial ports preventing the use of a $Ag/Ag^+$ reference electrode.

**Device Electrochemical Characterization**

Capacitor and hybrid capacitor devices were fabricated by directly printing on various substrates including, glass slides, ITO/Glass (Solaronix), and flexible ITO/PET (Aldrich).

The structures for devices were designed to either be symmetrical for top or bottom electrode in the case of the stacked configuration or to interlock in the case of the interdigitated electrode configuration. Two opposing electrodes were arranged on top of each other before sealing using commercial silicone sealant (Sherlock). The devices were allowed to cure for 24hrs before backfilling with electrolytes ($H_2SO_4$, or TEA $BF_4$/PC) and testing in a two-electrode configuration.

**ACKNOWLEDGEMENTS**

The authors would like to acknowledge Dr. Hisato Yamaguchi for insightful discussion. C.M. would like to acknowledge the EPSRC grants: EP/K01658X/1, EP/M022250/1, EP/K033840/1, EP/K016792/1, EP/K016407/1 and the EPSRC-Royal Society Fellowships Engagement Grant EP/L003481/1, and the award of a Royal Society University Research Fellowship by the UK Royal Society. P.C.S. would like to acknowledge funding through the H2020 Marie Sklodowska Curie Individual Fellowship (#660721). M.S.S. would like to acknowledge the financial support from the Imperial College President's PhD Scholarship scheme.

**Authors Contributions**

C.M. conceived the idea, C.M., P.C.S. and C.G. designed the experiments. C.G. performed the ink formulation, 3D printing, ink rheology, and mechanical testing. P.C.S. performed electrochemical, thermal, XRD, and SEM characterization, and data analysis. M.S.S. performed UV-Vis, SEM, and TEM experiments and data analysis. P.P. performed Raman spectroscopy and XPS data acquisition and analysis. K.S. performed electrochemical characterization. A.P. performed rheology measurements and analysis.